\documentclass[twocolumn,aps,pra]{revtex4}
\usepackage{graphicx}
\usepackage{amsmath}
\usepackage{tipa}
\usepackage{amssymb}

\begin{document}

\title{Separation induced resonances in quasi-one-dimensional ultracold atomic gases}
\author{Wenbo Fu,$^{1}$ Zhenhua Yu,$^{2}$ and Xiaoling Cui$^{1}$}
\email{xlcui@mail.tsinghua.edu.cn}
\affiliation{$^1$Institute for Advanced Study, Tsinghua University, Beijing, 100084 \\
$^2$Department of Physics, Ohio State University, Columbus, OH
43210}
\date{\small \today}
\begin{abstract}
We study the effective one-dimensional (1D) scattering of two
distinguishable atoms confined individually by {\it separated}
transverse harmonic traps. With equal trapping frequency for two
s-wave interacting atoms, we find that by tuning the trap
separations, the system can undergo {\it double} 1D scattering
resonance, named as the separation induced resonance(SIR), when the
ratio between the confinement length and s-wave
scattering length is within $(0.791,1.46]$. 
Near SIR, the scattering property shows unique dependence on the
resonance position. The universality of a many-body system on
scattering branch near SIR is demonstrated by studying the
interaction effect of a localized impurity coupled with a Fermi sea
of light atoms in a quasi-1D trap.

\end{abstract}

\maketitle

\section{Introduction}

Ultracold atomic gases have provided unprecedented accesses to
fascinating strongly interacting many-body systems, especially those
in unitary limit with resonant scattering. Besides Feshbach
resonances, various external confinements are recognized as another
efficient way to achieve resonances in all dimensions\cite{Olshanii,
Bergeman,Petrov, Peano, Nishida_mix}, and there have been successful
explorations of resonance scattering properties in
experiments\cite{Moritz05, Haller09, Lamporesi,Haller10,Kohl}.

The mechanism of confinement induced resonance was first explored by
Bergeman \emph{et al.}\cite{Bergeman}. A bound state constructed in
the Hilbert space spanned by the excited transverse states of
non-interacting Hamiltonian was introduced as a closed channel bound
state(CCBS). By tuning the confinements the resonance occurs when
CCBS touches the scattering threshold. This class of resonance can
also be understood as the consequence of modified low-energy
scattering theory by properly renormalizing all virtual scatterings
to high-energy states\cite{Cui}. The properties of these resonances
closely depend on the type of confinement potentials. Previous
studies have shown that within the contact interaction model, all
the induced resonances fall into two classes. If the trapping
potential decouples relative motion($r$) from center-of-mass($R$),
the only one CCBS would induce a {\it single} resonance such as in
Ref.\cite{Olshanii,Bergeman,Petrov}; if not, there would be an
infinite number of CCBS due to the coupling between $r$ and all
$R-$channels, resulting in an infinite number of resonances such as
in Ref.\cite{Peano,Nishida_mix}. 
Therefore an interesting question is whether these two classes have
covered all the possible resonances under external confinements. In
this paper, by providing an alternative class of scattering
resonance induced by trap separations(see below), we show the answer
is {\it no}.

We consider two distinguishable atoms confined individually by
transverse harmonic traps with tunable separations (see Fig.1(a)).
We find two resonances of one-dimensional(1D) scattering by tuning
the separation, which we name as ``separation induced resonance"
(SIR). It is the non-monotonic evolution of CCBS with the separation
that gives rise to the emergence of two resonances, and also leads
to new features in parameters describing the effective 1D
scattering. By introducing such a system, we show an interesting
class of induced resonances as SIR. For such resonances, the number
of resonances is not solely determined by the number of CCBS, and
the effective 1D scattering strength exhibits exotic dependence on
the tunable parameter as shown by Fig.3. All these features are
qualitatively different from those of Feshbach resonances and
previously studied confinement induced resonances. Our finding
substantially enriches the existing understanding of the induced
resonance physics. In addition, we study the many-body physics
across SIR. By employing an impurity problem, we show that a
many-body system on metastable scattering branch can go across
double SIRs and exhibit universal properties at each resonance. The
two-body bound state is also studied, and experimental realizations
and detections are discussed finally.

\begin{figure}[hbtp]
\includegraphics[width=85mm]{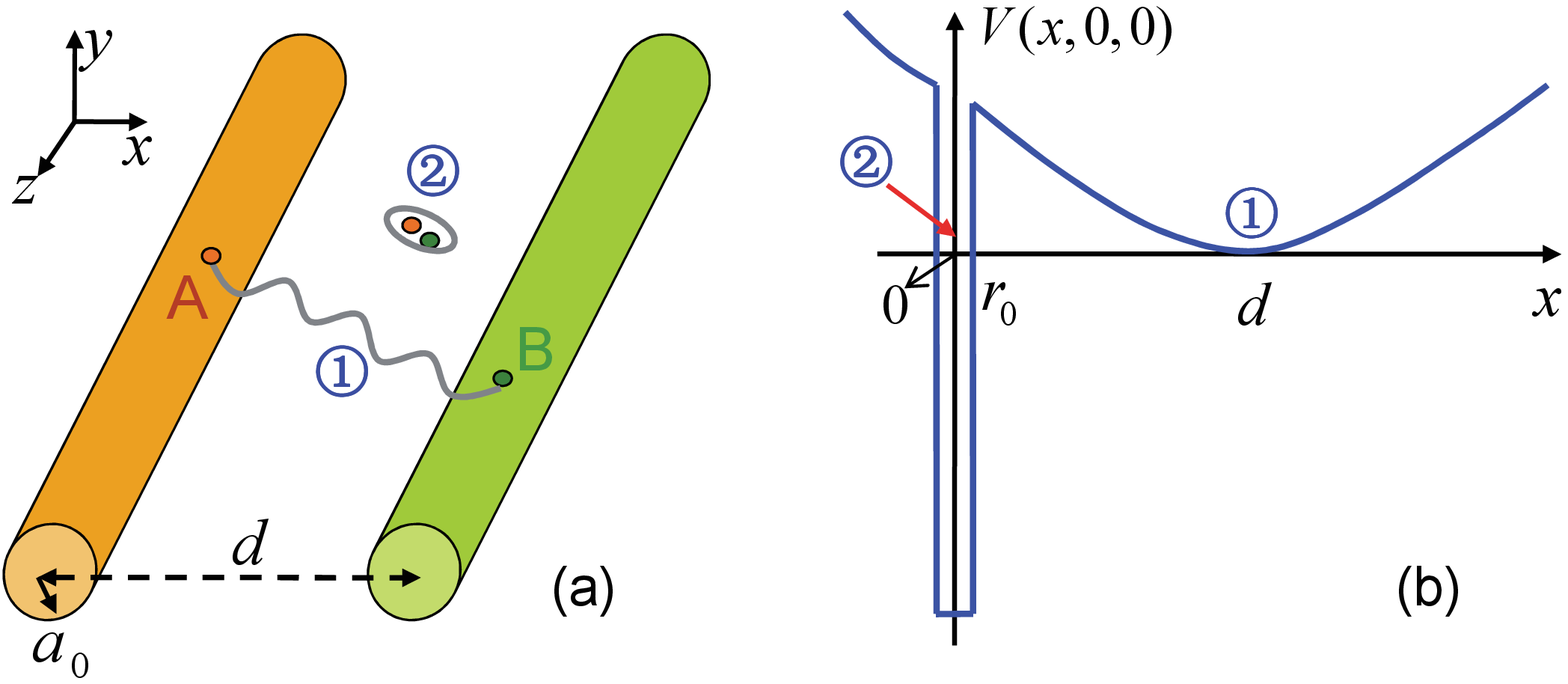}
\caption{(Color online) Schematic plot of system setup. (a)Two
interacting atoms(A, B) are separately confined in transverse
harmonic traps with characteristic length $a_0$ and distance $d$.
(b)Potentials $V(x,0,0)$ in the center-of-mass frame, including
trapping potential centered at $\mathbf{r}=(d,0,0)$(\textcircled{1})
and short-range($r_0\simeq0$) interaction at
$\mathbf{r}=0$(\textcircled{2}). } \label{fig1}
\end{figure}

The rest of the paper is organized as follows. In section II, we
present the formulism for effective 1D scattering across SIR. In
section III we discuss the origin of double resonances and show the
unique features of effective scattering strength near resonances.
The two-body bound state is studied in Section IV. Section V is
contributed to the impurity problem, from which we address the
universal property of a many-body system at metastable scattering
branch across SIR. We discuss the experimental realization and
finally remark on the generalization of SIR to other systems in
Section VI.

\section{Effective scattering in 1D}

We consider two distinguishable atoms, A and B, respectively trapped
by transverse potentials
$V_t(\mathbf{r}_A)=m_A\omega^2((x_A+d/2)^2+y_A^2)/2$ and
$V_t(\mathbf{r}_B)=m_B\omega^2((x_B-d/2)^2+y_B^2)/2$, which
decouples the center-of-mass and relative motions. The Hamiltonian
for the relative motion is $H_{rel}(\mathbf{r})=H_0+ U(\mathbf{r})$,
where (we take $\hbar=1$ throughout the paper)
\begin{equation}
H_0=-\frac{\nabla^2_{\mathbf{r}}}{2\mu}+\frac{\mu}{2}\omega^2((x-d)^2+y^2),
\end{equation}
$U(\mathbf{r})=\frac{2\pi
a_s}{\mu}\delta(\mathbf{r})\frac{\partial}{\partial
r}r|_{r\rightarrow0}$, with the reduced mass $\mu=m_Am_B/(m_A+m_B)$
and the s-wave scattering length $a_s$. $H_0$ determines the
spectrum as $E=E_{n_x,n_y}+k^2/(2\mu)$, with
$E_{n_x,n_y}=(n_x+n_y+1)\omega$ the eigen-energies for the
transverse eigen-states
$\phi_{n_x,n_y}(x,y)\equiv\psi_{n_x}(x-d)\psi_{n_y}(y),\ n_x,
n_y=0,1,2...$ ($\psi_n$ is the 1D harmonic oscillator
wavefunctions).

In this system, the low energy scattering processes for incoming
wave functions with $n_x=n_y=0$ and $k^2/2\mu\ll\omega$ can be
described by a 1D Hamiltonian for the relative motion (along
z-direction) as $h_{rel}=-(1/2\mu)(d^2/dz^2)+g_{1D}\delta(z)$, with
$g_{1D}$ the 1D coupling strength\cite{Olshanii}. To derive $g_{1D}$
in terms of $a_s$, we study the scattering wave function at low
energy $E=\omega+k^2/(2\mu)$,
\begin{equation}
\Psi(\mathbf{r})=\phi_{0,0}(x,y)e^{ikz}+fG(\mathbf{r},0),\label{Psi}
\end{equation}
where the Green function, $G(\mathbf{r},0)=\langle
\mathbf{r}|\frac{1}{E-\hat{H}_0+i\delta}|0\rangle$, is expressed as
\begin{equation}
G(\mathbf{r},0)=\sum_{n_x,n_y}\int_{-\infty}^{\infty}\frac{dp}{2\pi}
\frac{\phi_{n_x,n_y}(x,y)e^{ipz}\phi_{n_x,n_y}^{*}(0,0)}{E-E_{n_x,n_y}-p^2/(2\mu)+i\delta}.\label{Green}
\end{equation}
To solve the problem, it is essential to write the asymptotic form
of Eq.\ref{Green} at $\mathbf{r}\rightarrow 0$ as
$G(r,0)=-\frac{\mu}{2\pi r}+C(ka_0,\tilde{d})+o(r)$, where
$a_0=\sqrt{1/\mu\omega}$ is the characteristic length of the
harmonic oscillator, $\tilde{d}=d/a_0$. With the help of the
imaginary-time evolution operator of harmonic
oscillator\cite{Peano}, we obtain $C=\frac{\partial}{\partial
r}(rG(r,0))_{r\rightarrow 0}$ as
\begin{equation}
C(ka_0,\tilde{d})=\frac{\mu}{\pi
a_0}[\frac{1}{ika_0}e^{-\tilde{d}^2}-\frac{1}{\sqrt{2\pi}}A(ka_0,\tilde{d})],\label{C}
\end{equation}
and
\begin{equation}
A(ka_0,\tilde{d})=\int_0^{\infty}\frac{d\tau}{\sqrt{\tau}}e^{\frac{k^2a_0^2\tau}{2}}(
\frac{e^{-\tilde{d}^2\tanh\frac{\tau}{2}}}{1-e^{-2\tau}}-\frac{e^{\frac{-k^2a_0^2\tau}{2}}}{2\tau}-e^{-\tilde{d}^2}).\label{A}
\end{equation}
According to the Schrodinger equation $H_{rel}\Psi=E\Psi$, we find
the scattering amplitude as
\begin{equation}
f=\phi_{0,0}(0,0)[\frac{\mu}{2\pi a_s}-C(ka_0,\tilde{d})]^{-1},
\end{equation}
thus we obtain the closed form of Eq.\ref{Psi}. When $d=0$, all
equations reproduce the well-known quasi-1D results\cite{Olshanii,
Bergeman}. More detailed derivation of Eq.\ref{C} is given in
Appendix A.

At large distances along z-direction, all terms in the Green
function (Eq.\ref{Green}) decays except for the lowest transverse
mode with $n_x=n_y=0$\cite{Olshanii}. Since $U(\mathbf r)$ only
takes effect at $r=0$ but vanishes otherwise, the part of
$\Psi(\mathbf r)$ with even-parity of $z$, $\Psi_{\rm
even}(\mathbf{r})=\frac{1}{2}(\Psi(x,y,z)+\Psi(x,y,-z))$, is
scattered while the odd-parity part remains unaltered. By
Eq.~(\ref{Psi}), $\Psi_{\rm
even}(\mathbf{r})=\phi_{0,0}(x,y)e^{i\delta_k}\cos(k|z|+\delta_k)$,
where the phase shift $\delta_k$ satisfies
\begin{equation}
\tan\delta_{k}=-\frac{2}{k
a_0}e^{-\tilde{d}^2}[\frac{a_0}{a_s}+\sqrt{\frac{2}{\pi}}A(ka_0,\tilde{d})]^{-1}.
\label{delta}
\end{equation}
In the effective 1D limit($ka_0\ll 1$), we find
$A(ka_0,\tilde{d})=A(0, \tilde{d})+\lambda_d (ka_0)^2+o(k^4a_0^4)$,
where $0<\lambda_d\le0.666$ for all $\tilde{d}$ as shown by Fig.6 in
Appendix B. [See more details regarding the property of
$A(ka_0,\tilde{d})$ in Appendix B]. Therefore for low-energy
scattering, we neglect the energy-dependence in A-function and
approximate the coupling strength at different $k\ll 1/a_0$ to be a
constant given by $g_{1D}=-\lim_{k\rightarrow0}k\tan\delta_k/\mu$.
To this end, we obtain the relation between $g_{1D}$ and $a_s$.

According to Eq.\ref{delta}, the separation induced resonance (SIR)
for $k\to0$ occurs at
\begin{equation}
\frac{a_0}{a_s}=-\sqrt{\frac{2}{\pi}}A(0,\tilde{d}), \label{CIR}
\end{equation}
when $g_{1D}=\infty, \ |\delta|=\pi/2$. In Fig.2(a) we plot the
right-hand-side of Eq.\ref{CIR}, which is the required value of
$a_0/a_s$ for 1D scattering resonance at given $\tilde{d}$.
According to the properties of $A(0,\tilde{d})$ as given by
Eqs.(\ref{small},\ref{large}) in Appendix B, we obtain the
asymptotic expression of $a_0/a_s$ at SIR for small $\tilde{d}$ as
$[a_0/a_s]_{\rm SIR}\approx1.46-1.39\tilde{d}^2$, and for large
$\tilde{d}$ as $[a_0/a_s]_{\rm SIR}\approx\tilde{d}-1/\tilde{d}$. In
the limit of large $\tilde{d}$ and also large $a_0/a_s$, the
mechanism for SIR can be understood in the following intuitive way.
When a tightly bound molecule(play the role as CCBS\cite{Bergeman})
is formed, the relative wave function of the molecule concentrate
around $r=0$(shown as \textcircled{2} in Fig.1). Its energy is
approximately the sum of internal binding energy due to interaction,
$-1/2\mu a_s^2$, and external potential energy due to trap
separations, $\mu\omega^2d^2/2$; the resonance occurs when the total
energy matches the threshold of free particles(shown as
\textcircled{1} in Fig.1), i.e., $-1/2\mu
a_s^2+\mu\omega^2d^2/2=\omega$, which is exactly the criterion as
extracted from Eq.(\ref{CIR}) and Eq.(\ref{large}). For intermediate
$\tilde{d}$, there is a minimum of $[a_0/a_s]_{\rm SIR}$ as $0.791$
found at $\tilde{d}=1.123$. Remarkably, {\it for fixed $a_0/a_s \in
(0.791, 1.46]$, the system undergoes two distinctive resonances as
$\tilde{d}$ increases from zero.}

\begin{figure}[hbtp]
\includegraphics[height=5cm,width=8.5cm]{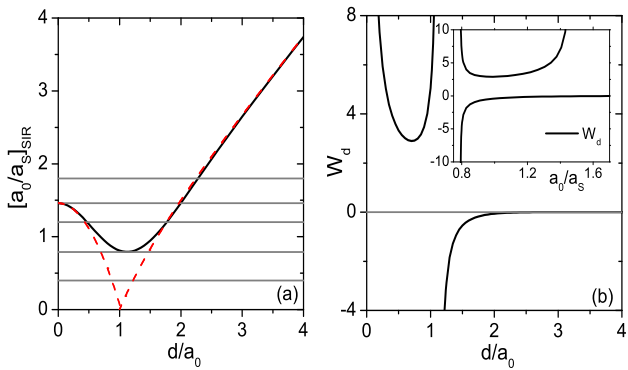}
\caption{Separation induced resonance(SIR) in quasi-1D system.
(a)$a_0/a_s$ as a function of $\tilde{d}=d/a_0$ at SIR. Dashed lines
are the functional fit to $1.46-1.39\tilde{d}^2$ in small
$\tilde{d}$, and $\tilde{d}-1/\tilde{d}$ in large $\tilde{d}$ limit.
The horizontal lines label five coupling strengths (from bottom to
top): $a_0/a_s=0.4,0.791,1.2,1.46,1.8$, with further properties
shown in Fig.3,4. (b)Resonance width $W_d$ versus resonance position
$\tilde{d}$ and corresponding $a_0/a_s$(inset).} \label{fig2}
\end{figure}

\section{Basic features of SIR}

In this section we address the basic features of SIR. First we
explore the origin of double resonances, and secondly we study the
resulted structure of effective scattering strength across SIR and
corresponding resonance width.

\subsection{Origin of double resonances}

The origin of double resonances can be understood by studying the
CCBS that is constructed by all excited transverse modes of $H_0$.
The binding energy of the CCBS, $E_b^c=E-\omega$, is given by
\begin{equation}
\frac{\mu}{2\pi a_s}=\lim_{r\rightarrow 0} (G^c(r,0)+\frac{\mu}{2\pi
r}),\label{ccbs}
\end{equation}
where $G^c$ is the closed channel Green function which follows
Eq.\ref{Green} but excludes $n_x=n_y=0$ in the summation. Thus
Eq.\ref{ccbs} is equivalent to
\begin{equation}
\frac{a_0}{a_s}=-\sqrt{\frac{2}{\pi}}A(\sqrt{\frac{2E_b^c}{\omega}},\tilde{d}),\label{ccbs_eq}
\end{equation}
Compared with Eq.~(\ref{CIR}), we see that SIR occurs when
$E_b^c=0$, i.e., when the CCBS touches the threshold.

\begin{figure}[hbtp]
\includegraphics[height=6cm,width=8.7cm]{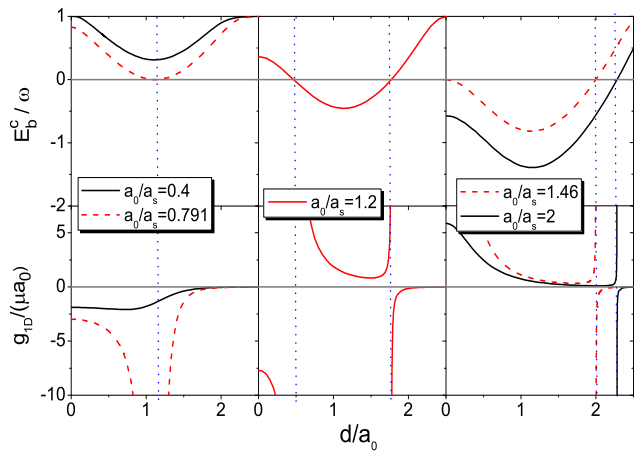}
\caption{The binding energy of CCBS($E_b^c/\omega$) and the
effective 1D coupling strength($g_{1D}/(\mu a_0)$) as functions of
$d/a_0$, for several typical values of $a_0/a_s$ that correspond to
five horizontal lines in Fig.2a. Vertical blue lines denote the
resonance positions $\tilde{d}_{res}$, where $E_b^c=0$ and
$g_{1D}=\infty$.} \label{fig3}
\end{figure}

The intriguing dependence of the resonance position on $\tilde{d}$
is a direct consequence of that of $E^c_b$ on $\tilde{d}$. In
Fig.\ref{fig3}(a), we show that for given $a_0/a_s$, $E_b^c$
decreases with $\tilde{d}$ at small $\tilde{d}$ but increases at
large $\tilde{d}$. Mathematically this is attributed to the
non-monotonic behavior of $A-$function(Eq.(\ref{A})) when increasing
$\tilde{d}$. For fixed $E_b^c$, A-function increases/decreases with
$\tilde{d}$ in the small/large $\tilde{d}$ limit (particularly these
properties are shown in Eqs.(\ref{small},\ref{large}) for
$A(0,\tilde{d})$); while for fixed $\tilde{d}$, A-function always
increases with the energy. These properties together give rise to
the non-trivial dependence of $E_b^c$ on $\tilde{d}$ as solved from
Eq.(\ref{ccbs_eq}) for each given $a_0/a_s$. Physically, the
non-trivial behavior of $E_b^c$ can be explained in the following
way. The explicit form of $G^c$ in Eq.~(\ref{ccbs})
(cf.~Eq.~(\ref{Green})) indicates that $d$ affects $E_b^c$ only
through the coupling weight
$\alpha_{n_x,n_y}=|\phi_{n_x,n_y}(0,0)|^2=\psi_{n_x}^2(-d)\psi^2_{n_y}(0)$,
particularly the $\psi_{n_x}^2(-d)$ part.
Unlike $d=0$ case where the interaction only couples even-parity
states ($n_x=0,2,...$), non-zero $d$ additionally mix all odd-parity
states ($n_x=1,3,...$) into $\Psi(\mathbf r)$. When $\tilde{d}\ll
1$, $\alpha_{n_x,n_y}$ increases with $\tilde{d}$ for all odd $n_x$,
while decreases for all even $n_x$. We find that the former effect
dominates so that small nonzero $\tilde d$ facilitates the formation
of CCBS and gives lower $E_b^c(d)\approx E_b^c(0)-3d^2/(2\mu
a_0^4)$. When $\tilde{d}\gg 1$, the coupling for both even and odd
$n_x$ decays exponentially with $\tilde{d}$. $E_b^c$ in this limit
approaches $\omega$ from below, implying two uncorrelated atoms when
trapped far apart.

As analyzed above, for any given $a_0/a_s$, $E^c_b$ as a function of
$\tilde{d}$ first decreases and then increases. As shown in Fig. 3,
for small $a_0/a_s<0.791$ the interaction is not strong enough to
make $E_b^c$ even touch the threshold and there is no resonance; for
large $a_0/a_s>1.46$, the interaction is so strong that at $d=0$ the
CCBS is already below the threshold, and therefore the resonance is
only possible at large $\tilde{d}$; however, for intermediate
$a_0/a_s \in (0.791, 1.46]$, $E^c_b$ is able to cross zero twice due
to its non-monotonic behavior, and correspondingly $g_{1D}$ diverges
whenever $E^c_b$ across zero. This is the origin of double resonance
feature.

\subsection{$g_{1D}$ near SIR}

Near SIR, $g_{1D}$ can be parameterized as $\tilde{g}_{1D}[\equiv
g_{1D}\mu a_0]=W_d/(\tilde{d}-\tilde{d}_{res})$, with
\begin{equation}
W_d=\sqrt{2\pi}e^{-\tilde{d}^2}[\frac{\partial
A(0,\tilde{d})}{\partial\tilde{d}}]_{res}^{-1}. \label{W}
\end{equation}
Here $W_d$ is the width of SIR, analogous to that defined in
Feshbach resonance\cite{Chin}.
In our system, $W_d$ reflects the coupling strength between the open
and closed channel responsible for SIR, and
also determines how well SIR can be accessed in experiment due to
the limited resolution of trap separations.

An important feature of SIR is that {\it the resonance width
strongly depends on the resonance position, which changes sign at
$\tilde{d}_{res}=1.123$}. We show in Fig.2(b) how $W_d$ changes with
$\tilde{d}_{res}$ and corresponding $a_0/a_s$. Due to the vanishing
$\frac{\partial A(0,\tilde{d})}{\partial\tilde{d}}$ around
$\tilde{d}_{res}=0$ and $1.123$, $W_d$ also experiences divergent
behavior asymptotically as $W_d\sim 1/d_{res}$ and $W_d\sim
-1/(d_{res}-1.123)$ respectively. At the left(right) side of
$d_{res}=1.123$, $W_d$ is positive(negative). Particularly, for a
given $a_0/a_s\in(0.791,1.46]$ the system accomplishes two SIRs: one
of them is with large positive $W_d$ and the other with small
negative $W_d$(see the inset of Fig.2(b)); the opposite signs of
these two $W_d$ determine that $g_{1D}$ keeps sign between two
SIRs(see Fig.3), in contrast with Feshbach resonance where $a_s$
crosses zero between two adjacent resonances.

In the limit of $a_0/a_s,\ \tilde{d}_{res}\gg 1$, $W_d$
exponentially decays as $W_d = -2e^{-\tilde{d}^2}$. Physically such
a narrow width corresponds to the very weak overlap of the
wavefunctions of interacting particles, i.e., A and B atoms have
little probability to collide with each other when they are trapped
far apart. Note that the width defined in Eq.\ref{W} is meaningful
for the realistic detection of SIR in experiment, but should not be
confused with the narrow width effect in a magnetic Feshbach
resonances\cite{Ho}. In fact, as shown in Appendix B, the
k-dependence of $g_{1D}$ is always very weak and even negligible in
$\tilde{d}\gg 1$ limit although the resonance width is exponentially
small.

\section{Two-body bound state}

The true two-body bound state is given by the pole of the scattering
amplitude, $f(i\kappa)=\infty$, i.e.,
\begin{equation}
\frac{\mu}{2\pi a_s}=C(i\kappa a_0,\tilde{d}).\label{Eb}
\end{equation}

\begin{figure}[hbtp]
\includegraphics[height=4.5cm, width=8.5cm]{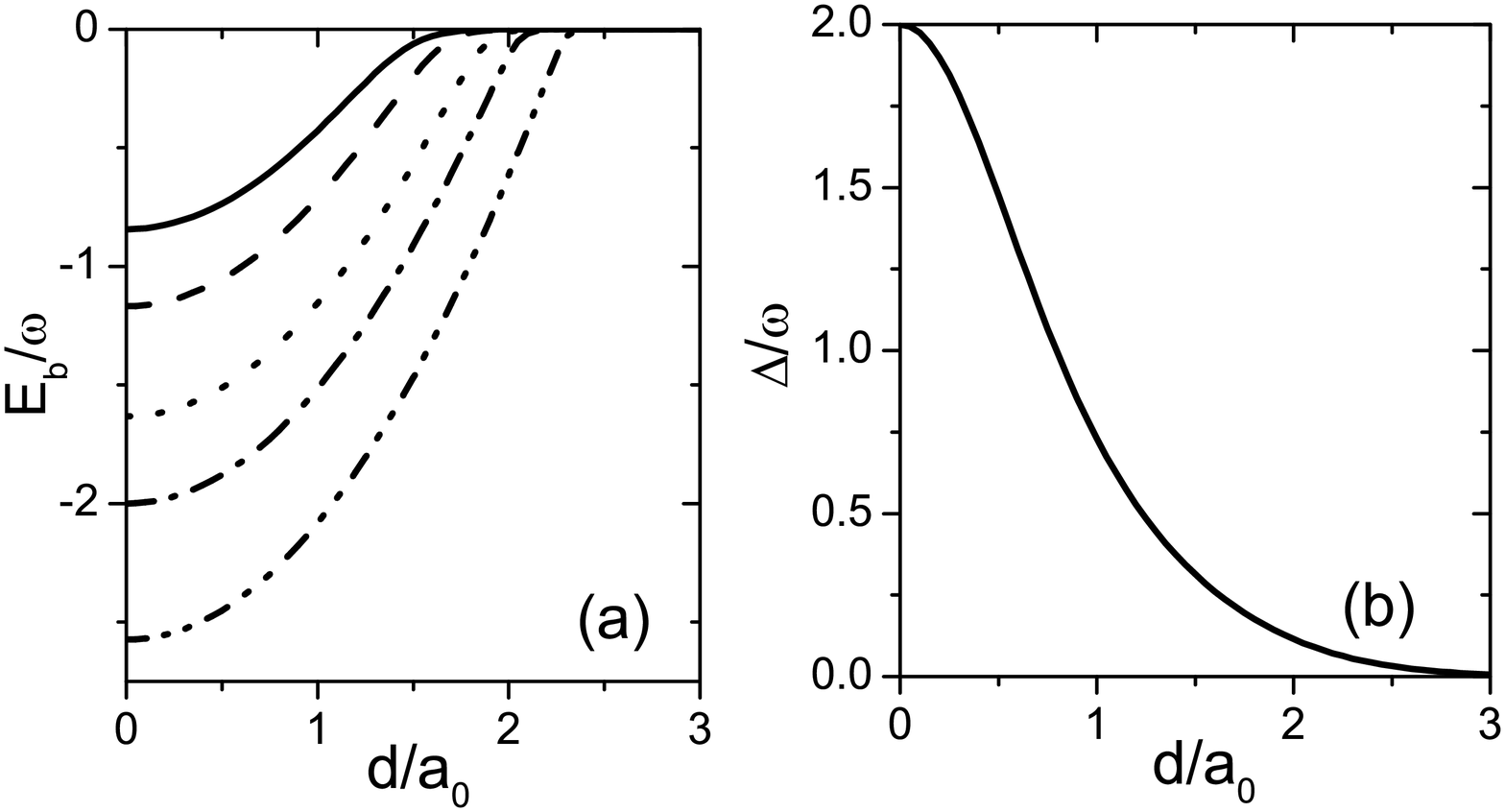}
\caption{(a)Binding energy of two-body bound state $E_b$ as a
function of $d/a_0$ for different
$a_0/a_s=0.4,0.791,1.2,1.46,1.8$(from top to bottom). (b) Energy
difference between CCBS and two-body bound state,
$\Delta=E_b^c-E_b$, right at SIR as a function of resonance position
$d/a_0$. } \label{fig_Eb}
\end{figure}

In Fig.\ref{fig_Eb}(a) the binding energy,
$E_b=E-\omega=-\kappa^2/(2\mu)$, is plotted as a function of
$\tilde{d}$. We see that $E_b$ always exists below zero for any
$a_s$ and $d$, due to the inclusion of $n_x=n_y=0$ mode and the
effective 1D density of state at low energies. Moreover, $E_b$
monotonically decreases with $\tilde{d}$, which is dramatically
different from $E_b^c$. In weak coupling limit($a_s/a_0\rightarrow
0^-$), the bound state is merely the 1D consequence which gives
$E_b=-\frac{\mu g_{1D}^2}{2}=-\frac{2a_s^2}{\mu
a_0^4}e^{-2\tilde{d}^2}$. In the strong coupling
limit($a_s/a_0\rightarrow 0^+$), for small $\tilde{d}$ it is
straightforward to obtain the binding energy as
$E_b(\tilde{d})=E_b(0)+\mu\omega^2d^2/2$, which is shifted up
exactly by the potential barrier; for large $\tilde{d}$, $E_b$ would
be very small and exponentially decay as $E_b=-2[\mu
a_0^2(\tilde{d}-a_0/a_s)^2]^{-1}e^{-2\tilde{d}^2}$. To explore the
difference between the CCBS and true two-body bound state, we plot
in Fig.\ref{fig_Eb}(b) their energy difference, $\Delta=E_b^c-E_b$,
right at SIR as a function of $\tilde{d}$. When $d=0$,
$\Delta=2\omega$ reproduces the result in Ref.\cite{Olshanii}. As
$\tilde{d}$ increases, $\Delta$ decreases and becomes exponentially
small in the large $\tilde{d}$ limit.

\section{Universality of scattering branch at SIR}

The universal property of a many-body system at SIR can be
effectively explored by considering the following exactly solvable
impurity problem. Consider that atoms A($m_A=m$) form a Fermi-sea
with Fermi energy $k_F$ and interact with a localized impurity
B($m_B=\infty$) at the origin. In this case the Hamiltonian for A is
exactly $H_{rel}$ with $\mu=m$ (cf. Fig.1(b)). The same phenomenon
of double SIRs can be deduced for A moving in the effective 1D tube.

Suppose the tube is with boundary $[-L,L]$, the allowed wave vectors
are given by $kL+\delta_k=(n+\frac{1}{2})\pi\ (n=0,1,...)$. Given
that there is no occupation of possible bound states, the non-zero
phase shifts $\delta_k$ give rise to the interaction energy (or the
energy difference from non-interacting case) as\cite{Combescot}
\begin{equation}
E_{int}=-\frac{1}{m\pi}\int_0^{k_F}k\delta_k dk.\label{Eint}
\end{equation}
Note that the scattering states contributing to $E_{int}$ here
belong to the metastable scattering branch in quasi-1D in which
$-\pi<\delta_k<0$ for all $k$. Explicitly, the metastable branch
corresponds to a Hilbert space expanded by scattering states, i.e.,
without the occupation of molecules. One typical example is the
repulsive Fermi gas in 3D with positive $a_s$, and recently there
have been extensive studies of single impurity problem in such
system both theoretically\cite{theory} and
experimentally\cite{expe}. Moreover, the scattering branch has also
been realized in the bosonic quasi-1D system\cite{Haller09} in the
absence of trap separations. In our system with trap separations, as
shown in Fig.\ref{fig_Eb}, a two-body bound state always exists
below the threshold for any $a_s$ and $\tilde{d}$, whose binding
energy monotonically decreases as $\tilde{d}$ increases. This
indicates that in a many-body system the universality is only
possible for the metastable scattering branch which excludes the
Hilbert space of molecules.

\begin{figure}[hbtp]
\includegraphics[height=4.6cm,width=8.1cm]{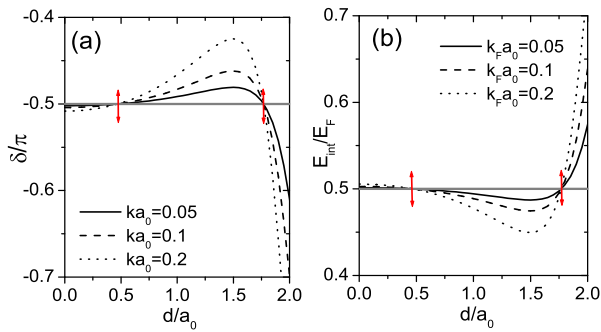}
\caption{Universality at SIR in an impurity system for given
$a_0/a_s=1.2$ ($a_0=1/\sqrt{m\omega}$). (a)$\delta_k$ in terms of
$\tilde{d}=d/a_0$ at given $ka_0=0.05,0.1,0.2$. The double SIR occur
at $\tilde{d}=0.47, 1.77$ (as shown by red arrows), where
$\delta_k=-\pi/2$(gray line) for any $k$. (b)$E_{int}/E_F$ as a
function of $\tilde{d}$ for different fermi momentum
$k_Fa_0=0.05,0.1,0.2$. $E_{int}$ shows universal value as
$E_F/2$(gray line) at SIR. } \label{fig4}
\end{figure}

According to Eq.\ref{delta}, in Fig.\ref{fig4}(a) we plot $\delta_k$
as a function of $\tilde{d}$ for different momenta $k$. Considering
$ka_0\ll1$, we have used $k$-independent $A(0,\tilde{d})$ in
Eq.\ref{delta}, which will bring a negligible correction of the
order of $o((ka_0)^2)$. We see that all the curves with different
$k$ cross exactly at the location of SIR. This is due to the {\it
universal} phase shift as $-\pi/2$ right at SIR for all values of
$k$, and according to Eq.\ref{Eint} this further leads to the {\it
universal} interaction energy as half of the Fermi energy,
$E_{int}=E_F/2$, at any position of SIR (see Fig.\ref{fig4}(b)).
According to Eq.\ref{delta} and Eq.\ref{W}, the slopes of $\delta$
and $E_{int}$ across SIR are both inversely proportional to the
resonance width $W_d$.

Above impurity problem explicitly shows that {\it the universal
behavior of a many-body system is a direct consequence of the
divergent two-body coupling strength and the resulted uniform phase
shift in k-space}. This conclusion should generally apply to a wide
range of cases, e.g., arbitrary numbers and mass ratios of
two-component fermions. In our system, we conclude that within
$a_0/a_s\in(0.791,1.46]$, by increasing $\tilde{d}$ the metastable
many-body scattering state would go across two consecutive universal
regimes. At small $\tilde{d}$ it is a crossover from the Fermionic
super-Tonks-Girardeau(TG)\cite{Chen} to Fermionic TG regime, and at
large $\tilde{d}$ a reverse process.

\section{Experimental realization and final remarks}

The probing of SIR and its physical consequences can be realized by
taking advantage of sophisticated optical techniques to manipulate
cold atoms. For two spin-species of the same isotope ($m_A=m_B$),
separated harmonic traps can be generated in the setup of
spin-dependent optical lattices\cite{Bloch, Mckay}. The separation
$d$ can be tuned by adjusting the polarization angles of two
linearly polarized and counterpropagating laser beams(with
wavelength $\lambda$), which create the lattices. In
Ref.\cite{Bloch}, the maximum separation between the nearest two
species is $d_{max}=3\lambda/16\sim150nm$, comparing to the typical
confinement length $a_0\sim50nm$. The ratio, $d/a_0=0\sim3$, is of
most interest as shown by Fig.2 and Fig.3. For different isotopes
$m_A\neq m_B$, the separated traps can be achieved by fine tuning
the laser frequency according to different atomic transition lines
for different atoms\cite{Lamporesi, Spethmann}. In such cold atom
systems, the position of SIR can be pinned down by the maximum of
atom loss rate\cite{Lamporesi,Haller10, Haller09}; $g_{1D}$ can be
mapped out from the frequencies of collective modes\cite{Haller09};
the binding energy or interaction energy can be deduced from the
shift of peak frequency of atomic transition using rf
spectroscopy\cite{Kohl}.

Before closing, we emphasize that SIR should fall into a new class
of resonance different from magnetic Feshbach resonances and
confinement induced resonances. The effect brought about by the trap
separation is so generic that it should also hold for other trap
geometries or interaction types. Our further studies find that the
basic features of SIR, i.e., the non-monotonic evolution of CCBS
with the separation and thus the resulted exotic properties of
effective scattering strength near resonance, still persist in
quasi-2D geometry or in the presence of p-wave
interaction\cite{more}. Even in the case of $\omega_A\neq \omega_B$
in our quasi-1D system, where the center-of-mass and relative
motions can not be separated, each CCBS emerging from different
center-of-mass channels will evolve non-monotonically with
$\tilde{d}$. It is conceivable that for given $a_s/a_0$ several CCBS
could go across the zero threshold energy. This will bring extra
interesting resonance properties, such as arbitrarily finite
resonance number and novel structure of $g_{1D}$ near resonance. In
contrast, for Feshbach resonance or confinement induced resonances,
each CCBS monotonically evolves with the magnetic field or
confinement length, resulting in infinite number of resonances and
the similar structure of effective scattering strength near each
resonance\cite{Peano,Nishida_mix,Lamporesi}. The SIR found at the
two-body level also has strong indication for intriguing many-body
phenomena, which await full exploration in the future.

We thank Hui Zhai for helpful discussions and comments on the
manuscript. Authors XC and ZY thank the Aspen Center for Physics for
its hospitality during the cold atom workshop. This work is
supported in part by Tsinghua University Initiative Scientific
Research Program, NSFC under Grant No. 11104158, NSF Grants
DMR-0907366, and by DARPA under the Army Research Office Grant
Nos.~W911NF0710464, W911NF0710576.

\appendix

\section{Derivation of Eq.\ref{C}}

We expand the Green function (Eq.(3) in the text) as
\begin{widetext}
\begin{multline}
G(r,0)=\phi_{0,0}(x,y)\phi_{0,0}^{*}(0,0)e^{ik|z|}\frac{\mu}{ik}-\int_0^{\infty}dt\sqrt{\frac{\mu}{2\pi t}}e^{-\frac{\mu z^2}{2t}+\frac{k^2t}{2\mu}}\\
((\sum_{n_1=0}^{\infty}e^{-n_1\omega
t}\phi_{n_1}(x)\phi_{n_1}^{*}(0))(\sum_{n_2=0}^{\infty}e^{-n_2\omega
t}\phi_{n_2}(y)\phi_{n_2}^{*}(0))
-\phi_{0,0}(x,y)\phi_{0,0}^{*}(0,0)),\label{G1d}
\end{multline}
\end{widetext}
where we have used imaginary-time integration for the low-energy
scattering ($E=\omega+\frac{k^2}{2\mu}<2\omega$); $\mu$ is the
reduced mass of two atoms A and B; $\phi_n(x)=\psi_n(x-d)$,
$\phi_n(y)=\psi_n(y)$, and
$\psi_n(x)=\frac{1}{\sqrt{\mathcal{N}}}H_n(\frac{x}{a_0})\exp(-\frac{x^2}{2a_0^2})$
is the eigen-state for 1D harmonic oscillator centered at $x=0$ and
with characteristic length $a_0$.

\begin{widetext}
Further by utilizing the single-particle imaginary-time propagator
($a_0=1/\sqrt{\mu\omega}$)
\begin{equation}
\sum_{n=0}^{\infty}e^{-n\omega t}\psi_{n}(x)\psi_{n}^{*}(x')
=\frac{1}{\sqrt{\pi}a_0}\frac{1}{\sqrt{1-e^{-2\omega
t}}}e^{-\frac{x^2+x'^2}{2a_0^2}\coth(\omega
t)+\frac{xx'}{a_0^2\sinh(\omega t)}},
\end{equation}
the second term in Eq.\ref{G1d} is reduced to
\begin{equation}
-\frac{1}{\pi a_0^2}\sqrt{\frac{\mu}{2\pi
w}}\int_0^{\infty}\frac{d\tau}{\sqrt{\tau}}e^{\frac{k^2a_0^2\tau}{2}-\frac{z^2}{2a_0^2\tau}}(\frac{e^{-\frac{(x-d)^2+(-d)^2}{2a_0^2}\coth\tau+\frac{(x-d)(-d)}{a_0^2\sinh\tau}-\frac{y^2}{2a_0^2}\coth\tau}}{1-e^{-2\tau}}-e^{-\frac{(x-d)^2}{2a_0^2}-\frac{d^2}{2a_0^2}-\frac{y^2}{2a_0^2}}),
\end{equation}
\end{widetext}
which implicitly includes a divergence as $-\frac{\mu}{2\pi r}$ at
small $r=\sqrt{x^2+y^2+z^2}$ due to the integration at
$\tau\rightarrow 0$. Using the exact relation
\begin{equation}
\int_{0}^{\infty}\frac{d\tau}{\tau^{\frac{3}{2}}}e^{-\frac{r^2}{2a_0^2\tau}}=\sqrt{2\pi}\frac{a_0}{r}
\end{equation}
we extract the divergent term from the Green function and rewrite it
as
\begin{widetext}
\begin{multline}
G(r,0)=\phi_{0,0}(x,y)\phi_{0,0}^{*}(0,0)e^{ik|z|}\frac{\mu}{ik}-\frac{1}{\pi a_0^2}\frac{\mu a_0}{\sqrt{2\pi}}\int_0^{\infty}\frac{d\tau}{\sqrt{\tau}}e^{\frac{k^2a_0^2\tau}{2}-\frac{z^2}{2a_0^2\tau}}\\
(e^{-\frac{(x-d)^2+(-d)^2}{2a_0^2}\coth\tau+\frac{(x-d)(-d)}
{a_0^2\sinh\tau}-\frac{y^2}{2a_0^2}\coth\tau}\frac{1}{1-e^{-2\tau}}-\frac{e^{-\frac{x^2+y^2}{2a_0^2\tau}-\frac{k^2a_0^2\tau}{2}}}{2\tau}
-e^{-\frac{(x-d)^2}{2a_0^2}-\frac{d^2}{2a_0^2}-\frac{y^2}{2a_0^2}})-\frac{\mu}{2\pi
r}.
\end{multline}
\end{widetext}
Now it is straightforward to obtain
$C(ka_0,\tilde{d})=\frac{\partial}{\partial
r}(rG(r,0))|_{r\rightarrow 0}$ as Eq.(4) and further
$A(ka_0,\tilde{d})$ as Eq.(5) in the text.

\section{Property of $A(ka_0,\tilde{d})$}

To see the energy-dependence of $A(ka_0,\tilde{d})$, we expand it in
terms of small $ka_0\ll 1$ as $A(ka_0,\tilde{d})=A(0,
\tilde{d})+\lambda_d (ka_0)^2+o(k^4a_0^4)$, with
\begin{eqnarray}
\lambda_d&=&\int_0^{\infty}d\tau\frac{\sqrt{\tau}}{2}(\frac{e^{-\tilde{d}^2\tanh
\frac{\tau}{2}}}{1-e^{-2\tau}}-e^{-\tilde{d}^2})\nonumber\\
&=&\left\{\begin{array}{l}
0.409,\ \ d=0 \\
\frac{\sqrt{2\pi}}{4\tilde{d}},\ \ \tilde{d}\rightarrow\infty
\end{array}\right. \label{k-depend}
\end{eqnarray}
Note that due to $\partial \lambda_d/\partial \tilde{d}>0$ at
$\tilde{d}\ll 1$, $\lambda_d$ increases with $\tilde{d}$ at small
$\tilde{d}$. However, $\lambda_d$ decrease as $1/\tilde{d}$ at large
$\tilde{d}$. Therefore in the intermediate $\tilde{d}$, $\lambda_d$
should reach a maximum determined by $\partial \lambda_d/\partial
\tilde{d}=0$, i.e.,
\begin{equation}
\int_0^{\infty}d\tau\frac{\sqrt{\tau}}{2}(\frac{e^{-\tilde{d}^2\tanh\frac{\tau}{2}}\tanh\frac{\tau}{2}}{1-e^{-2\tau}}-e^{-\tilde{d}^2})=0,
\end{equation}
and this gives $\tilde{d}=1.023$ and $(\lambda_d)_{max}= 0.666$.

\begin{figure}[hbtp]
\includegraphics[height=5cm,width=7cm]{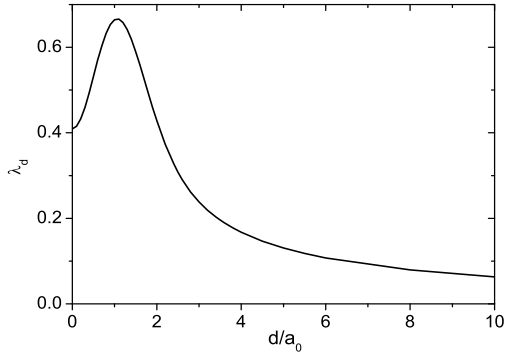}
\caption{$\lambda_d$(Eq.\ref{k-depend}) as functions of the
separation $\tilde{d}$. } \label{fig_supple}
\end{figure}

The small $\lambda_d$ obtained for all $\tilde{d}$ justify us to
only consider the $k-$independent part of $A(ka_0,\tilde{d})$ in the
scattering problem,
\begin{equation}
A(0,\tilde{d})=\int_0^{\infty}\frac{d\tau}{\sqrt{\tau}}(
\frac{e^{-\tilde{d}^2\tanh\frac{\tau}{2}}}{1-e^{-2\tau}}-\frac{1}{2\tau}-e^{-\tilde{d}^2}).\label{A_d}
\end{equation}

Next we analyze it in two limits.

(1) when $\tilde{d}\rightarrow 0$,
$A(0,\tilde{d})=a+b\tilde{d}^2+o(\tilde{d}^4)$, with
\begin{eqnarray}
a&=&\int_0^{\infty}\frac{d\tau}{\sqrt{\tau}}(\frac{1}{1-e^{-2\tau}}-\frac{1}{2\tau}-1)\simeq-1.83,\nonumber\\
b&=&\int_0^{\infty}\frac{d\tau}{\sqrt{\tau}}(1-\frac{\tanh\frac{\tau}{2}}{1-e^{-2\tau}})\simeq1.75.
\label{small}
\end{eqnarray}
This gives the resonance value of
$[a_0/a_s]|_{res}=1.46-1.39\tilde{d}^2$ at small $\tilde{d}$.

(2) when $\tilde{d}\rightarrow \infty$, Eq.\ref{A_d} is mostly
contributed by small $\tau$. In this limit, Eq.\ref{A_d} is
equivalent to
\begin{eqnarray}
A(0,\tilde{d})&\approx&\int_0^{\infty}\frac{d\tau}{\sqrt{\tau}}(
\frac{e^{-\tilde{d}^2\tau/2}}{2\tau(1-\tau)}-\frac{1}{2\tau})\nonumber\\
&\approx&
\int_0^{\infty}[\frac{e^{-\tilde{d}^2\tau/2}-1}{2\tau^{3/2}}+
\frac{e^{-\tilde{d}^2\tau/2}}{2\tau^{1/2}}]\nonumber\\
&=&\sqrt{\frac{\pi}{2}}(-\tilde{d}+1/\tilde{d}),\label{large}
\end{eqnarray}
which gives $[a_0/a_s]|_{res}=\tilde{d}-1/\tilde{d}$ at large
$\tilde{d}$.

Eq.\ref{small} and Eq.\ref{large} show that $A(0,\tilde{d})$
increases with $\tilde{d}$ at small $\tilde{d}$ while decreases at
large $\tilde{d}$. The turning point is given by $\frac{\partial
A(0,\tilde{d})}{\partial \tilde{d}}=0$, i.e.,
\begin{equation}
\int_0^{\infty}\frac{d\tau}{\sqrt{\tau}}(\frac{e^{-\tilde{d}^2\tanh\frac{\tau}{2}}\tanh\frac{\tau}{2}}{1-e^{-2\tau}}-e^{-\tilde{d}^2})=0,
\end{equation}
and this gives $\tilde{d}=1.123$, corresponding to the resonance
value of $[a_0/a_s]|_{res}=0.791$ (see also Fig.2(a)).

\end{document}